\documentclass[12pt]{article}
\usepackage{epsf}
\usepackage[dvips]{graphicx}
\usepackage{amssymb}

\begin{document}
\title{Total cross section of neutron-proton scattering at low energies in quark-gluon model}

\author{{\slshape V. A.  Abramovsky\footnote{Victor.Abramovsky@novsu.ru}, N. V. Radchenko\footnote{Natalya.Radchenko@novsu.ru}}\\[1ex]
Novgorod State University, 173003 Novgorod the Great, Russia}
\date{}
\maketitle

\begin{abstract}
We show that analysis of nonrelativistic neutron-proton scattering
in a framework of relativistic QCD based quark model can give
important information about QCD vacuum structure. In this model we
describe  total cross section of neutron-proton scattering at
kinetic energies of projectile neutron from 1~eV up to 1~MeV.
\end{abstract}

1. Numerous experimental data are collected in low energy nuclear
physics, substantial part of which demands rethinking in a
framework of modern understanding of strong interaction physics.
In particular, in wide range of nonrelativistic kinetic energies
of projectile neutron the  total cross section of neutron-proton
scattering is constant and is approximately equal to 20 barn
(\cite{bib1}, see also http://www.nndc.bnl.gov/sigma/).

The theoretical explanation of the  total cross section
constancy~\cite{bib2} is based on $S$-wave scattering where total
cross section is given by the following formula
\begin{equation}\label{1}
\sigma_{tot}^{(t)}=\frac{4\pi\hbar^2}{M}\frac{1}{E+E_d},
\end{equation}
where $M$ -- nucleon mass, the deuteron binding energy
$E_d\simeq2.23$~MeV. The upper index in~(\ref{1}) means that the
scattering takes place in triplet state. However, if one
substitutes all numerical values into~(\ref{1}) (which has no free
parameters), one obtains $\sigma_{tot}^{(t)}\simeq2$ barn at
energies $E\ll E_d$, this is one order of magnitude smaller than
the experimental value. So in order to describe the experimental
data one has to introduce contribution from neutron-proton system
singlet state with an unclear phenomenological parameter $E_s$ --
energy of this system virtual state .

2. Quantum chromodynamics is a microtheory of strong interactions.
QCD self-consistently describes processes in the region of
``hard'' physics, where running coupling constant $\alpha_S$ is
small and perturbation theory can be applied. In the region of
``soft'' physics, where distances are large and transferred
momenta are small, there is no justification for QCD diagrams
applicability. But, using our previous experience of strong
interactions analysis in colorless particles language, where the
Feynman diagrams allowed to explain inelastic processes structure
at high energies, we will consider QCD diagrams in this region.

3. The example of QCD Feynman  diagrams successful use in ``soft''
physics is the description of hadron-hadron interactions cross
sections in wide range of ultrarelativistic
energies~\cite{Radchenko:2010mw}. In this work we have
successfully described total cross sections of $pp$, $\bar{p}p$,
$np$, $\bar{p}n$, $\pi^\pm p$, $K^\pm p$ scatterings by the same
parameters in the framework of one approach. The main idea of this
approach lies in fact that constant part of raising cross sections
is described by one gluon exchange between components with only
valence quarks (Fig.~1). The convolution of this diagram with its
adjoint diagram (Fig.~2a) gives contribution to the total cross
section of nucleon-nucleon scattering. Since color objects do not
fly out, the $F_n$ amplitudes correspond to processes which take
place with probability equals to 1. Therefore the diagram from
Fig.~2a is equivalent~\cite{Radchenko:2010mw} to diagram of the
Low-Nussinov two-gluon pomeron~\cite{Low:1975sv}. Namely such
diagram gives constant contribution to total cross section.
\begin{figure}[!h]
\centerline{
\includegraphics[scale=0.55]{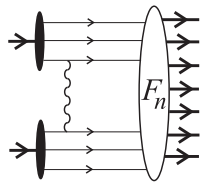}}
\caption{ One-gluon exchange (wavy line) between hadrons
components containing only valence quarks (thin solid lines).
Thick solid lines correspond to hadrons both in initial and final
states.}
\end{figure}
\begin{figure}[!h]
\centerline{
\includegraphics[scale=0.55]{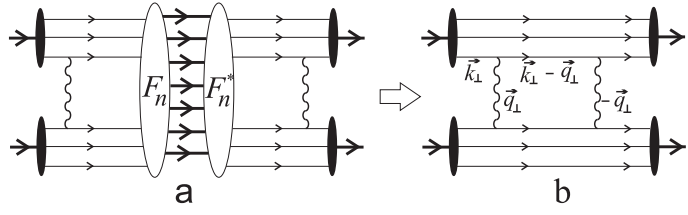}}
\caption{ a) convolution of one-gluon exchange diagram with its
adjoint diagram; b) two-gluon exchange diagram -- Low-Nussinov
two-gluon pomeron.}
\end{figure}

4. The diagram of two-gluon pomeron for meson-meson scattering was
calculated in~\cite{Gunion:1976iy} with $\pi$-meson model form
factor as in~\cite{Feynman:1973xc}. Constant contribution to
$\pi\pi$ scattering is equal to
\begin{equation}\label{2}
\sigma_{tot}^{\pi\pi(const)}=\frac{128}{9}\,\pi\,\frac{\alpha_S^2}{m_\rho^2},
\end{equation}
where $m_\rho$ -- $\rho$-meson mass. Using Levin-Frankfurt
relations of total cross sections ratios for hadron-hadron
scattering~\cite{Levin:1965mi}
\begin{equation}\label{3}
\sigma_{tot}^{\pi\pi}:\sigma_{tot}^{\pi
p}:\sigma_{tot}^{pp}=4:6:9,
\end{equation}
the last two are in good agreement with experiment, we obtain the
expression for constant contribution to neutron-proton scattering
at high energies
\begin{equation}\label{4}
\sigma_{tot}^{np(const)}=32\,\pi\,\frac{\alpha_S^2}{m_\rho^2}.
\end{equation}
Given the value of $pp$, $\bar{p}p$, $np$, $\bar{p}n$ total cross
sections constant part $\sigma_0=21.63$~\cite{Radchenko:2010mw} we
can substitute it in~(\ref{4}) and thus we obtain
\begin{equation}\label{5}
\alpha_S=0.576,\quad\alpha_S^2=0.332.
\end{equation}

5. The two-gluon exchange amplitude (Fig.~2b), which is defined in
ultrarelativistic region, must also exist for nonrelativistic
energies.

First, since non particle degrees of freedom (e.g. instantons)
exist in QCD, particle degrees of freedom -- quarks and gluons
also must exist. Therefore two-gluon diagrams, generally speaking,
must give contribution also in nonrelativistic region.

Second, in ultrarelativistic case of ``soft'' scattering, where
the most part of hadron-hadron total cross section is gathered, no
noticeable contributions to interaction through non particle
degrees of freedom are seen experimentally. Also, non particle
degrees of freedom effects are not taken into account in various
fits of total cross sections. (We think that non particle degrees
of freedom affect QCD vacuum.)

Third, the constant contribution to total cross sections is
provided only by vector exchange, i.e. only two-gluon diagram can
give constant value of neutron-proton scattering.

Therefore we accept that two-gluon diagram gives contribution in
nonrelativistic region.

6. Unlike the ultrarelativistic case, where all valence quarks in
the upper block of Fig.~1 move in one direction and in the lower
block in another (we consider scattering in center-of-mass frame),
in the nonrelativistic case the situation is different. We deal
with almost rest neutron and proton. We also accept that valence
quarks in neutron and proton have current masses of a few MeV. If
one supposes that in a rest hadron quarks momenta are of the same
magnitude as transverse momenta of quarks in the ultrarelativistic
case, since transverse part of momentum does not change under
Lorentz transformation, then the value of rest hadron quark
momentum is about 0.5 -- 0.6~GeV/c. In this case we have to
consider relativistic model of quarks interaction even in rest
hadron.

7. It is easy to show that additional contribution exists in
almost rest neutron and proton, which is equal to contribution
from the two-gluon diagram in the ultrarelativistic region. It is
connected to movement of quarks in the upper block of diagram in
Fig.~1 in negative direction while quarks in the lower block move
in positive direction. This contribution decreases like $4M^2/s$
as the energy increases. Then the two-gluon exchange contribution
to the amplitude of elastic scattering in nonrelativistic region
has the following form
\begin{equation}\label{6}
\sigma_{tot}^{np}=32\frac{\alpha_S^2}{m_\rho^2}\left(1+\frac{4M^2}{s}\right).
\end{equation}

8. We accept the hypothesis of ``freezing'' of running constant
$\alpha_S$~\cite{Dokshitzer:1998nz} -- \cite{Simonov:2001dt} and
assume that for nonrelativistic neutron-proton scattering
$\alpha_S$ is equal to the value, obtained from total cross
sections fittings in the ultrarelativistic region,
$\alpha_S=0.576$. Substituting this value in~(\ref{6}) we obtain
in the nonrelativistic case $\sigma_{tot}^{np}\simeq43$~mb. This
is 50 times lower than the experimental value of
$\sigma_{tot}^{np}$. It is possible to assume that the value of
$\alpha_S$ is large in the region of nonrelativistic scattering,
we will consider this question in future works. But here we
suppose that the running constant $\alpha_S$ is freezed and thus
we find a physical mechanism that increases total cross section
$\sigma_{tot}^{np}$.

9. In vacuum quark fluctuations light quarks with sufficiently
small momenta $k<\Lambda$ move on distances larger than the
confinement radius. In this case color electric field string
appears between them. If quarks did not carry spin, tension of
string would rotate quarks in direction to each other not allowing
them to move further than the confinement radius. But quarks are
fermions. Therefore, in order to change the momentum direction,
either chirality must be changed or it is needed to rotate quark
spin. So Casher~\cite{Casher:1979vw} has proposed that quark
fluctuations in vacuum are not isolated. Thus it is possible to
replace quarks from correlated fluctuations instead of rotating
spin. In hadrons $q\bar{q}$, $qqq$ all the time valence quarks
come closer to hadron boundary they are replaced by quarks from
quark condensate. There is no such effect in the ultrarelativistic
case since contribution from  vacuum fluctuations can be neglected
there.

\begin{figure}[!h]
\centerline{
\includegraphics[scale=0.55]{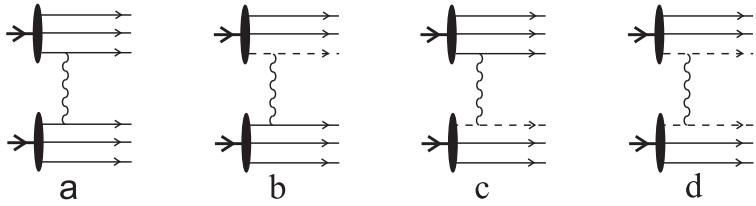}}
\caption{ a) diagram of one-gluon exchange between valence quarks
in hadrons; b), c) diagrams of one-gluon exchange between valence
quark in one of hadrons and quark from quark-gluon condensate
which is replacing valence quark in another hadron; d) diagram of
one-gluon exchange between quarks from quark-gluon condensate
replacing valence quarks in both hadrons.}
\end{figure}

10. At low kinetic energies of colliding neutron and proton there
are no secondary particles coming from color strings decay.
Therefore one-gluon exchange amplitude is unlike the diagram in
Fig.~1 and is shown in Fig.~3a. Its modulus squared directly gives
two-gluon diagram from Fig.~2b. We take into account the effect of
valence quark replacement with a quark from quark condensate by
considering diagrams in Fig.~3b, 3c. Presumably, interference
between quark on mass shell in hadron and virtual quark from
quark-gluon condensate is not possible. Therefore we take moduli
squared of diagrams in Fig.~3b, 3c. Diagram in Fig.~3d, describing
interaction of quarks from condensate, must be neglected.

The interaction radius of quarks in nucleon with quarks from
vacuum condensate is defined by the correlation length, i.e. the
distance, where quarks from vacuum fluctuations can replace quarks
in nucleons. We define this distance as $R_{cor}=1/m_G$.
Therefore, by analogy with the ultrarelativistic case we write
down the phenomenological formula
\begin{equation}\label{7}
\sigma_{tot}^{np}=32\,\pi\,\alpha_S^2(m_\rho^2)\left[\frac{1}{m_\rho^2}\left(1+\frac{2M}{E_{kin}^n
M}\right)+\frac{2}{m_G^2}\frac{\sqrt{2}\,m_\Lambda}{\sqrt{E_{kin}^nM+2m_\Lambda^2}}\right].
\end{equation}
We explain formula~(\ref{7}). The first term in square bracket is
evident, it follows from~(\ref{6}). Factor $2/m_G^2$ in the second
term gives contribution from interaction of valence quark from one
of the nucleons with quark from vacuum fluctuation, replacing
valence quark. Evidently, vacuum fluctuations contribution has to
decrease as energy of projectile neutron increases (i.e. square of
total energy in center-of-mass system increases). It is taken into
account by factor
$\sqrt{2}\,m_\Lambda/\sqrt{E_{kin}^nM+2m_\Lambda^2}$ in the second
term.

11. We fitted the experimental data on neutron-proton total cross
section~\cite{bib1} in kinetic energy range from 1~eV to 1~MeV
using formula~(\ref{7}) and obtained values of $m_G=35$~MeV and
$m_\Lambda=4.7$~MeV, the result is shown in Fig.~4.
\begin{figure}[!h]
\centerline{
\includegraphics[scale=0.58]{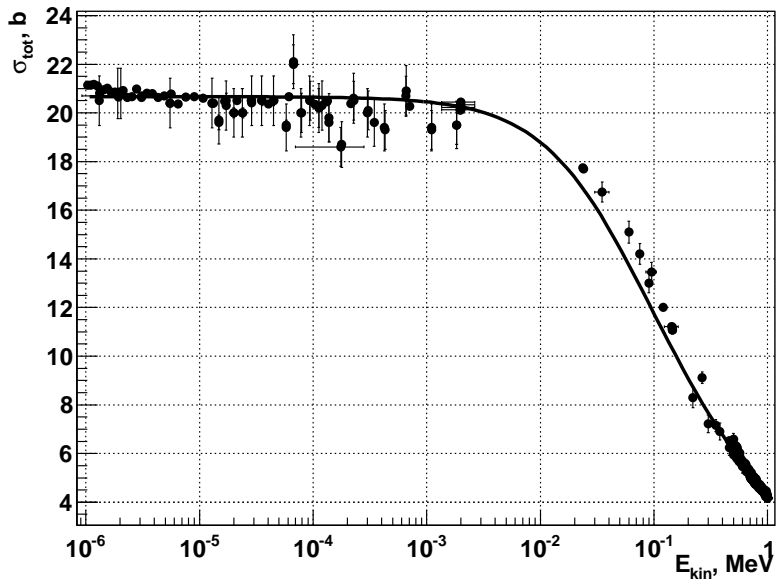}}
\caption{ Fitting of $\sigma_{tot}^{np}$ by formula~(\ref{7}).}
\end{figure}

12. In conclusion we repeat the logical scheme of our approach.
\begin{enumerate}
    \item We suppose that quarks are relativistic in rest hadrons.
    \item We accept that the QCD Feynman diagrams can be used for
    phenomenological description at low nonrelativistic energies
    of colliding nucleons.
    \item The constant contribution to total cross section of
    neutron-proton scattering comes from the two-gluon exchange in
    imaginary part of elastic scattering amplitude.
    \item There is effect of infrared ``freezing'' of the running
    coupling constant $\alpha_S$ and its value can be prolonged in
    region of low nonrelativistic energies.
    \item At low energies there is replacement of valence quarks
    in nucleons by quarks from vacuum fluctuations -- Casher
    mechanism.
\end{enumerate}

With these assumptions it is possible to describe the value of
neutron-proton scattering total cross section in neutron energy
range from 1~eV to 1~MeV.

In our opinion, the proposed phenomenological model has right to
exist and it is not worse than description of cross section
$\sigma^{np}$ which is given in nuclear theory. In future works we
will give more detailed justification of our assumptions, based on
experimental data.

\vspace{1cm} \textbf{Acknowledgement}

One of authors (N.V. Radchenko) gratefully acknowledges financial
support by grant of Ministry of education and science of the
Russian Federation, federal target program ``Scientific and
scientific-pedagogical personnel of innovative Russia'' 2009 --
2013, grant P1200.


\begin{thebibliography}{99}
\bibitem{bib1}
M.~B.~Chadwick et al.  Nuclear Data Sheets. {\bf 107}, 2931
(2006).

\bibitem{bib2}
  Yu.~M.~Shirokov and N.~P.~Yudin
  ``Nuclear physics,''
{\it  Imported Pubn 1983.}


\bibitem{Radchenko:2010mw}
  N.~V.~Radchenko and A.~V.~Dmitriev,
  arXiv:1010.5259 [hep-ph].


\bibitem{Low:1975sv}
  F.~E.~Low,
  Phys.\ Rev.\  D {\bf 12}, 163 (1975).

  S.~Nussinov,
  Phys.\ Rev.\ Lett.\  {\bf 34}, 1286 (1975).


\bibitem{Gunion:1976iy}
  J.~F.~Gunion and D.~E.~Soper,
  Phys.\ Rev.\  D {\bf 15}, 2617 (1977).

  E.~M.~Levin and M.~G.~Ryskin,
  Sov.\ J.\ Nucl.\ Phys.\  {\bf 34} (1981) 619
  [Yad.\ Fiz.\  {\bf 34} (1981) 1114].


\bibitem{Feynman:1973xc}
  R.~P.~Feynman,
  ``Photon-hadron interactions,''
{\it  Reading 1972, 282p}


\bibitem{Levin:1965mi}
  E.~M.~Levin and L.~L.~Frankfurt,
  JETP Lett.\  {\bf 2}, 65 (1965).


\bibitem{Dokshitzer:1998nz}
  Y.~L.~Dokshitzer,
  arXiv:hep-ph/9812252.


\bibitem{Badalian:1999fq}
  A.~M.~Badalian and V.~L.~Morgunov,
  Phys.\ Rev.\  D {\bf 60}, 116008 (1999)
  [arXiv:hep-ph/9901430].

\bibitem{Simonov:2001dt}
  Yu.~A.~Simonov,
  Phys.\ Atom.\ Nucl.\  {\bf 66}, 764 (2003)
  [Yad.\ Fiz.\  {\bf 66}, 796 (2003)]
  [arXiv:hep-ph/0109159].

\bibitem{Casher:1979vw}
  A.~Casher,
  Phys.\ Lett.\  B {\bf 83}, 395 (1979).


\end{thebibliography}
\end{document}